\documentclass[11pt, a4paper]{article}
\usepackage[a4paper, margin=2.5cm]{geometry}
\usepackage{amsmath,amssymb,amsthm,mathtools}
\newtheorem{theorem}{Theorem}[section]
\newtheorem{corollary}[theorem]{Corollary}
\newtheorem{proposition}[theorem]{Proposition}
\theoremstyle{definition}\newtheorem{assumption}{Assumption}
\usepackage{xcolor}
\definecolor{accent}{RGB}{24, 64, 112}
\usepackage[colorlinks=true, linkcolor=accent, urlcolor=accent, citecolor=accent]{hyperref}

\title{\textbf{Comment on ``An algebraic quantization of gravity from the null cone''}}
\author{Christopher J. Mogni}
\date{November 29, 2026}

\begin{document}

\maketitle

\begin{abstract}
In \emph{An algebraic quantization of gravity from the null cone}, Chandrasekaran argues that capping a gravitational clock selects a nontrivial, finite-dimensional Hilbert space for a causal diamond. We show that neither existence of continuum states below the cap nor finite dimensionality of the accepted space is established. On a round background of fixed positive area, the lower edge of the regulated spectrum of the positive, unsubtracted linearized clock tends to infinity as radial resolution is refined. At fixed local area density, classical non-linear transport preserves the leading short-distance singularity of the shear bracket. Finite positive focusing requires the shear commutator expectation to tend to zero on shrinking radial packets. Whether the coupled quantum theory satisfies this condition remains open. Even granting existence and an area ceiling, the gluing argument requires independent control of convergence: identity gluing yields a finite squared norm if and only if the accepted space is finite-dimensional. Finally, for a lower-bounded vacuum-subtracted quadratic clock obeying standard displacement identities, the spectral projection onto $(-\infty,\ell]$ has infinite rank for every $\ell$ strictly above the spectral infimum. The packets used in this proof have divergent exact classical clock readings at fixed nonzero amplitude as they approach the corner.
\end{abstract}

\section{Introduction}

Black hole thermodynamics and the covariant entropy bound~\cite{Bousso} suggest that a bounded region of spacetime holds only finitely many independent quantum states, controlled by the area of its boundary. Deriving such a count directly from the local degrees of freedom of gravity remains an open problem. In continuum quantum field theory on a fixed background, local observable algebras are generally of type~III, admitting no trace and no density matrices. Semi-classical gravitational constructions can improve this to type~II algebras, which admit a trace, but they still lack minimal projections and retain infinitely many degrees of freedom~\cite{Witten,CLPW}. Deriving a finite-dimensional Hilbert space from the dynamics of four-dimensional general relativity, without imposing a cutoff by hand, would give gravitational entropy a microscopic foundation.

In \emph{An algebraic quantization of gravity from the null cone}~\cite{Ven}, Chandrasekaran proposes such a derivation for a finite causal diamond. He solves the gravitational constraints on its intersecting null boundaries, quantizes the resulting radiative and corner data at finite $G_N$, and defines a gravitational clock from the affine time elapsed along the incoming null generators. For fixed initial area element and expansion, gravitational shear delays a generator from reaching a prescribed area element at the corner. Because shear carries the two graviton polarizations in four dimensions, capping the elapsed affine time restricts both the corner area and the gravitational radiation. To show that this cap selects a nonzero, finite-dimensional Hilbert space, from which a discrete area spectrum and a finite Lorentzian thermal trace follow, Chandrasekaran's argument requires two independent steps: first, establishing that some finite clock bound leaves a nonzero space of states in the continuum limit; and second, using the resulting ceiling on the corner area, together with a gluing argument across a corner boost, to rule out infinitely many orthogonal states.

In this comment, we demonstrate that neither step is established in the continuum limit and that the gravitational clock faces a structural tension. To preserve the classical focusing inequality that caps the corner area, Chandrasekaran orders the squared shear as a positive operator, retaining its quantum zero-point fluctuations. On a round background of fixed positive area, the uncertainty relation for the linearized shear on short intervals of a single generator forces the minimum possible focusing to grow in every state as those intervals shrink. Even with fixed angular resolution, the minimum clock reading in every state eventually exceeds any finite bound as radial resolution is refined (Corollary~\ref{cor:clock}). The existence estimate in Appendix~C.2 of~\cite{Ven} is carried out at fixed finite resolution, where this limit is never taken.

The same reasoning yields a necessary condition for any non-linear quantization. On a state with finite positive focusing, the shear commutator expectation on localized packets must vanish as their radial supports shrink. At fixed area density, classical parallel transport of the two polarizations preserves the leading singularity of the linearized bracket. Whether it survives quantization when the corner data fluctuate remains an open obstruction (Section~\ref{sec:non-linear}).

Subtracting the vacuum contribution avoids this zero-point divergence, but it changes the premise of the construction, because a renormalized shear square can take negative values~\cite{WallShear}, and the area ceiling would require a new lower bound. Moreover, for a lower-bounded vacuum-subtracted quadratic clock, gravitational shear localized very close to the corner barely delays the arrival of a generator while still carrying independent quantum fluctuations. Under the hypotheses of Theorem~\ref{thm:subtracted-clock}, the spectral projection onto $(-\infty, \ell]$ has infinite rank for every threshold $\ell$ strictly above the minimum possible reading, coinciding with $[0, \ell]$ if the clock is nonnegative, though the same perturbations produce divergent exact classical readings at fixed amplitude.

Even if one grants that the clock selects a nonempty space of states with bounded corner area, the gluing proof of finite dimension still requires independent justification. Section~5.2 of~\cite{Ven} infers finite dimension by comparing an amplitude on the uncut diamond with a sum over intermediate states after cutting at the corner. Once the corner area is bounded, however, every intermediate state enters the identity closure with a weight bounded away from zero. The squared norm of the closed preparation is finite if and only if the space being counted is already finite-dimensional. For the signed observable used in the boost argument, the step requiring justification is therefore the convergence of the cut sums to the prescribed uncut amplitude; Section~\ref{sec:hidden} shows how a single normalized probability on the uncut geometry could provide that independent input.

The paper is organized as follows. Section~\ref{sec:setup} reviews the gravitational clock, its positive ordering, and the two-step structure of Chandrasekaran's proof. Section~\ref{sec:weyl} proves the linearized continuum obstruction to existence, and Section~\ref{sec:non-linear} derives the local cancellation condition required of any non-linear quantization (with the classical parallel-transport estimates in Appendix~\ref{sec:geometry}). Section~\ref{sec:hidden} analyzes the closed norm and normalized gluing in the dimension argument. Section~\ref{sec:subtracted-clock} proves infinite rank for the vacuum-subtracted quadratic clock and evaluates the exact classical reading on its packets, while Appendix~\ref{sec:repairs} examines normal ordering and angular smearing of the focusing operator itself. Section~\ref{sec:discussion} summarizes the implications and states a conjecture for the full coupled shear bracket. Equation and section numbers attributed to the original paper refer throughout to~\cite{Ven}.

\section{The clock and its positive focusing form}\label{sec:setup}

To test whether a bound on elapsed affine time can select a nonzero, finite-dimensional Hilbert space, we first review how Chandrasekaran's gravitational clock links radiative shear to corner area, and how its operator ordering allows us to test the clock using a single graviton polarization. Physically, gravitational shear slows the expansion of a null congruence. At fixed initial area element and expansion, a sheared incoming null generator therefore takes more affine time to reach a prescribed area element at the corner than a shear-free generator. Averaging this elapsed affine time over the corner defines the clock reading, which constrains both the corner geometry and the radiation on the cone.

Consider a finite causal diamond $\mathcal{D}$ bounded by an incoming null cone $\mathcal{H}_p$ and an outgoing null cone $\mathcal{H}_q$, with small spacelike caps $\Sigma_0, \Sigma_1$ excising the past and future vertices and a 2-sphere corner $S = \mathcal{H}_p \cap \mathcal{H}_q$. On each cut of $\mathcal{H}_p$, we write the area element as $dA = \varrho^2 \epsilon_0$, where $\epsilon_0$ is the unit-sphere area form and $\varrho > 0$ is the local areal radius. Denoting its value at the corner by $\varrho_S(y)$ for each null generator $y \in S$, the total corner area is $\mathcal{A} = \int_S \varrho_S(y)^2 \epsilon_0$, and the corresponding dimensionless area is $s = \mathcal{A}/(4G_N\hbar)$.

Along each generator $y \in S$, the areal coordinate $v = \varrho/\varrho_S(y)$ runs from a fixed cutoff $v = \epsilon \in (0,1)$ at $\partial\Sigma_0$ to $v = 1$ at $S$. The two graviton polarizations are encoded in the unit-determinant conformal metric $\bar{q}_{AB}$ of the cuts, which Chandrasekaran parametrizes by two real fields $z^1, z^2$ with field-space metric $\operatorname{diag}(1, e^{2z^1})$ (Eq.~(4.21)). We write $w$ for the radial integration variable denoted by $s$ in~\cite{Ven}, because $s$ already denotes the dimensionless area. Integrating the null Raychaudhuri constraint from the past cap ($v = \epsilon$) to $v$ yields the nonnegative focusing integral of Eq.~(3.26c),
\begin{equation}
\Xi_y(v) = \frac{1}{8}\int_\epsilon^v w\,(\partial_w \bar{q})^{AB}\partial_w \bar{q}_{AB}\,dw = \frac{1}{4}\int_\epsilon^v w\left[(\partial_w z^1)^2 + e^{2z^1}(\partial_w z^2)^2\right] dw \ge 0.
\label{eq:classical_xi}
\end{equation}
The factor $e^{\Xi_y(v)}$ enters the conversion from areal coordinate to affine parameter, and equals one in the shear-free case. Multiplying the elapsed affine parameter $u(1,y) - u(\epsilon,y)$ by the initial expansion rate $p_{\epsilon,y} = (\partial_u\varrho)_{\epsilon,y}$ makes the local clock invariant under generator-dependent affine rescalings. Integrating along each generator and averaging over the corner yields the local and averaged classical clock readings (Eqs.~(3.27)--(3.28)),
\begin{equation}
u_*(y) = p_{\epsilon,y}\bigl[u(1,y) - u(\epsilon,y)\bigr] = \varrho_S(y)\int_\epsilon^1 e^{\Xi_y(v)}\,dv, \qquad \bar{u}_* = \frac{\int_S \varrho_S(y)\,u_*(y)\,\epsilon_0}{\int_S \varrho_S(y)\,\epsilon_0}.
\label{eq:classical_clock}
\end{equation}

In the quantum theory of~\cite{Ven}, the incoming radiative and corner canonical pairs at fixed total area $\mathcal{A}$ are represented on a continuum bosonic Fock space $\mathcal{K}_{\mathcal{A}}$, built on a product Gaussian reference state $|\Omega_{\mathcal{A}}\rangle$ over all radial and angular modes together with the corner data (Section~4.2). A pure extension of this state to the full observable algebra yields a Gelfand--Naimark--Segal (GNS) Hilbert space $\mathcal{H}_\omega$ that allows total-area fluctuations and carries an infinite-dimensional type~I factor (c.f. Eq.~(4.17)). Even at fixed $\mathcal{A}$, the local area density $\hat{\varrho}_S(y)$ can fluctuate across $S$. In Eqs.~(4.26)--(4.27), the quantum clock $\hat{\bar{u}}_*$ commutes with the total area operator $\hat{\mathcal{A}}$ and is defined on $\mathcal{K}_{\mathcal{A}}$ by the quadratic form
\begin{equation}
\big\|\hat{\bar{u}}_*^{1/2}|\psi\rangle\big\|^2 = \int_S \epsilon_0(y)\int_\epsilon^1 dv\, \big\| e^{\hat{\Xi}_y(v)/2}\,\hat{\varrho}_S(y)\,\hat{a}^{-1/2}|\psi\rangle\big\|^2, \qquad \hat{a} = \int_S \hat{\varrho}_S\,\epsilon_0.
\label{eq:clock_form}
\end{equation}
Imposing a maximum diamond size $\ell(\mathcal{D})$ restricts $\mathcal{H}_\omega$ to the projected subspace $\mathcal{H}_{\mathcal{D}} = P_{\mathcal{D}}\mathcal{H}_\omega$, where $P_{\mathcal{D}} = \mathbf{1}_{[0,\ell(\mathcal{D})]}(\hat{\bar{u}}_*)$ is the spectral projection onto clock readings at most $\ell(\mathcal{D})$ (Eqs.~(5.1)--(5.2)). We call a normalized state $\psi$ \emph{accepted} if $P_{\mathcal{D}}\psi = \psi$. Whether the infinite-dimensional type~I factor reduces to a finite-dimensional matrix algebra depends entirely on the rank of $P_{\mathcal{D}}$.

Chandrasekaran obtains the stated ceiling on the corner area from non-negativity of the focusing operator: $\hat{\Xi}_y(v) \ge 0$ implies $e^{\hat{\Xi}_y(v)} \ge \mathbf{1}$, and combining the area normalization $\int_S \hat{\varrho}_S(y)^2 \epsilon_0 = \hat{\mathcal{A}}$ with the Cauchy--Schwarz inequality on $S$ bounds the clock from below at fixed area. On the projected space $\mathcal{H}_{\mathcal{D}}$, the corner area therefore obeys (Eqs.~(5.21d) and (5.30a))
\begin{equation}
0 \le P_{\mathcal{D}}\hat{\mathcal{A}}P_{\mathcal{D}} \le \frac{4\pi \ell(\mathcal{D})^2}{(1-\epsilon)^2}P_{\mathcal{D}}, \quad \text{or, equivalently,} \quad 0 \le \hat{s} \le s_* \equiv \frac{\pi \ell(\mathcal{D})^2}{G_N\hbar(1-\epsilon)^2}.
\label{eq:area_bound}
\end{equation}
To preserve this classical sign after quantization, Eq.~(4.22) orders the focusing operator as
\begin{equation}
\hat{\Xi}_y(v) = \frac{1}{4}\int_\epsilon^v w\left[(\partial_w \hat{z}^1)^\dagger (\partial_w \hat{z}^1) + (\partial_w \hat{z}^2)^\dagger e^{2\hat{z}^1} \partial_w \hat{z}^2\right] dw,
\label{eq:quantum_xi}
\end{equation}
where $\hat{z}^\alpha = \mathcal{Q}(z^\alpha)$ denotes the Weyl quantization of $z^\alpha$ in~\cite{Ven}. The two terms are ordered differently. At linear order, $(\partial_w \hat{z}^1)^\dagger (\partial_w \hat{z}^1)$ coincides with the Weyl quantization of $(\partial_w z^1)^2$, whereas the second term is ordered as $A^\dagger B A$ with $B = e^{2\hat{z}^1} > 0$, ensuring positivity where ordinary Weyl quantization need not preserve sign. At any finite resolution, dropping the nonnegative second term leaves the form bound
\begin{equation}
\hat{\Xi}_y(v) \ge \frac{1}{4}\int_\epsilon^v w\,(\partial_w \hat{z}^1)^\dagger (\partial_w \hat{z}^1)\, dw.
\label{eq:z1_bound}
\end{equation}
Any scalar lower bound $b\mathbf{1}$ on the first polarization's contribution is therefore a lower bound on $\hat{\Xi}_y(v)$, and by the spectral theorem gives $e^{\hat{\Xi}_y(v)} \ge e^b \mathbf{1}$. This allows us to test the clock using a single polarization.

With the clock and its positive ordering in place, the finiteness proof in Section~5 of~\cite{Ven} proceeds in two stages. First, Section~5.1 argues that the incoming Gaussian reference state $|\Omega_{\mathcal{A}}\rangle$ has nonzero projection onto the allowed range of clock readings, $P_{\mathcal{D}}|\Omega_{\mathcal{A}}\rangle \neq 0$, using a regularized resolvent state to sidestep a possibly divergent vacuum expectation value of $e^{\hat{\Xi}_y(v)}$. Second, Section~5.2 uses the area ceiling~\eqref{eq:area_bound} to argue by contradiction that $\dim \mathcal{H}_{\mathcal{D}} < \infty$. If infinitely many orthogonal states existed, infinitely many would crowd into an arbitrarily narrow band of corner areas, where a relative corner boost would act nearly as a constant phase and fail to reproduce the boost dependence of a two-sided graviton correlator on the uncut diamond. We examine the first stage in Sections~\ref{sec:weyl} and~\ref{sec:non-linear} and the second stage in Section~\ref{sec:hidden}.

\section{A no-go theorem for positive continuum focusing}\label{sec:weyl}

We begin with the first stage of Chandrasekaran's argument, which seeks to show that some finite clock cap $\ell(\mathcal{D})$ admits a nonzero state. To test this condition cleanly, we hold the corner geometry fixed on a spherical, shear-free background of positive area $\mathcal{A} > 0$. Because the classical shear vanishes on this background, any lower bound on the focusing comes entirely from quantizing the radiation. The core physical mechanism is an uncertainty bound along a single null generator. Keeping the integrated shear square small requires fluctuations in every pair of orthogonal radial profiles to be simultaneously small, whereas for the packet pairs constructed below, the uncertainty relation forces the product of their uncertainties (and hence their squared norms) to grow as the packets narrow.

In this section, we prove that no nonzero state has finite unsubtracted focusing in the linearized continuum limit (Theorem~\ref{thm:linear}) and then show that this drives the entire regulated clock spectrum past every finite reading, even at fixed angular resolution (Corollary~\ref{cor:clock}).

Fix a resolved null generator $y \in S$ and an endpoint $v \in (\epsilon, 1)$. For any real test function $f$ supported in $(\epsilon, v)$, define the radially smeared shear of the first polarization by
\[
  \mathsf{S}(f) = \int_\epsilon^v f(w)\sqrt{w}\,\partial_w\hat{z}^1(w)\,dw.
\]
Let $q_\psi$ denote the expectation value of the unsubtracted focusing form~\eqref{eq:quantum_xi} on a normalized vector $\psi$ in its candidate form domain. By the single-polarization bound~\eqref{eq:z1_bound} and Bessel's inequality on $L^2(\epsilon, v)$, for every finite orthonormal family $\{f_i\}$ of real test functions supported in $(\epsilon, v)$,
\begin{equation}
  4q_\psi \ge \sum_i \|\mathsf{S}(f_i)\psi\|^2.
  \label{eq:bessel}
\end{equation}
The integral of the squared norm is the supremum of the right-hand side over all such finite families. Commutators below are understood as sesquilinear forms on the common field domain, so only a single application of each smeared field to $\psi$ is required.

\begin{theorem}[Positive focusing for the linearized null brackets]\label{thm:linear}
Fix $v > \epsilon$. Suppose the bracket of one resolved polarization on an open subinterval of $(\epsilon, v)$ has sign-kernel coefficient $-\gamma_y/(wt)$ with $\gamma_y > 0$ as in Eq.~(A.36) of~\cite{Ven}, and a smooth boundary kernel there. In any regular representation realizing these smeared commutators as forms on the candidate form domain, the unsubtracted focusing form has domain $\{0\}$. In fact, no nonzero vector satisfies~\eqref{eq:bessel} with finite $q_\psi$.
\end{theorem}

\begin{proof}
On a fixed generator, the linearized bracket takes the form
\[
  [\hat{z}^1(w), \hat{z}^1(t)]/i
  = -\gamma_y\,\frac{\operatorname{sgn}(w-t)}{wt} + k_\partial(w,t),
  \qquad \gamma_y > 0,
\]
where the constant $\gamma_y$ absorbs $\hbar$ and the angular normalization, and $k_\partial(w,t)$ is the boundary kernel fixed by the cut conditions. In the linearized fixed-density bracket, the homogeneous radial solutions are proportional to $1/w$, so $k_\partial$ is smooth on any compact interior subinterval of $(\epsilon, v)$. Differentiating both arguments and multiplying by $\sqrt{wt}$ yields a leading singularity $2\gamma_y(wt)^{-1/2}\delta'(w-t)$; the remaining terms have at most a delta singularity and bounded coefficients on compact interior intervals.

The $\delta'$ singularity pairs one radial profile with the derivative of the other, so its expectation on unit-norm profiles of width $h$ grows as $h^{-1}$, whereas the smooth boundary kernel remains bounded on interior tests. To isolate this effect, choose a real unit-norm bump function $F \in C_c^\infty(-1,1)$ and set $G = F'/\|F'\|_2$. Then $F$ and $G$ are orthonormal in $L^2(\mathbb{R})$ and satisfy $\int F(x)G'(x)\,dx = -\|F'\|_2 \neq 0$. Centering the rescaled unit-norm profiles $F_h(w) = h^{-1/2}F((w-w_0)/h)$ and $G_h(w) = h^{-1/2}G((w-w_0)/h)$ at an interior point $w_0 \in (\epsilon, v)$ and pairing them with $\delta'$ gives, up to orientation,
\begin{equation}
  [\mathsf{S}(F_h), \mathsf{S}(G_h)]/i
  = \frac{2\gamma_y}{w_0 h}\int F(x)G'(x)\,dx + O(1).
  \label{eq:packet}
\end{equation}
Here, the derivative acting on the smooth coefficient $(wt)^{-1/2}$ produces an $O(1)$ term, any delta term is bounded by Cauchy--Schwarz, and any bounded kernel contributes $O(h)$ because $\|F_h\|_1, \|G_h\|_1 = O(\sqrt{h})$. These estimates are uniform for $w_0$ in any fixed compact interior subinterval.

For every normalized vector $\psi$ satisfying~\eqref{eq:bessel}, the Robertson uncertainty relation~\cite{Robertson} gives
\begin{equation}
  4q_\psi \ge \|\mathsf{S}(F_h)\psi\|^2 + \|\mathsf{S}(G_h)\psi\|^2
  \ge \big|\langle [\mathsf{S}(F_h), \mathsf{S}(G_h)] \rangle_\psi\big|.
  \label{eq:robertson}
\end{equation}
By~\eqref{eq:packet}, the right-hand side diverges as $h \to 0$, so no nonzero vector has finite focusing. Because this lower bound depends only on the smeared commutators, it is state-independent and cannot be removed by changing the reference vacuum while preserving the smeared commutators.
\end{proof}

One shrinking pair already establishes Theorem~\ref{thm:linear}. To obtain a quantitative lower bound for the regulated clock spectrum below and for the non-linear comparison in Appendix~\ref{app:many}, we place many such pairs side-by-side in a fixed interior interval below $v_0 < 1$. For even $N$, divide the interval into $N/2$ equal cells and place one rescaled pair $(X_k, Y_k)$ in each cell, with width $h$ a fixed fraction of the cell length. All $N$ test functions are orthonormal. Orienting each pair so that its linearized commutator is $i\lambda_k$ with $\lambda_k > 0$, Eq.~\eqref{eq:packet} yields, for $N$ sufficiently large,
\begin{equation}
  \Lambda_N \equiv \sum_{k \le N/2} \lambda_k \ge c_1 \gamma_y N^2,
  \label{eq:regulated_floor}
\end{equation}
where the constant $c_1 > 0$ depends only on the chosen interval and the profiles $F, G$. Shortening the packets both increases each commutator as $N$ and allows $N/2$ disjoint pairs to fit in the interval, which produces the $N^2$ growth. This bound uses a fixed family of smooth radial tests, requires no assumption about the ordering of a mode basis, and needs no commutators between distinct pairs. For a regulator defined by a mode expansion, one first resolves the finite set of smooth tests at fixed $N$ so that their smeared brackets converge, and only then increases $N$.

Because the clock~\eqref{eq:clock_form} averages over the corner $S$, we must also check that angular averaging cannot dilute this radial lower bound. On a spherical background, every retained angular harmonic shares the same radial bracket up to its angular weight, so rotational invariance makes the coefficient $\gamma_y$ uniform over $S$. Specifically, for a rotation-invariant angular cutoff $J \ge 2$, the spin-weighted addition theorem~\cite{Goldberg} gives $A_J(y) = \sum_{j=2}^J (2j+1)/(4\pi)$, and the coefficient in~\eqref{eq:packet} is $\gamma_y = \gamma_J^{\rm harm} \propto A_J(y)/s$, independent of $y$. Throughout, $\gamma_y$ without a superscript denotes the positive reference coefficient used in the packet calculation; its lattice normalization is given separately in~\eqref{eq:A36}. The radial obstruction already holds at any such fixed $J$.

\subsection{Existence of states below the clock bound}\label{sec:structure}

It remains to pass from focusing on a single generator to the spectrum of the corner-averaged clock $\hat{\bar{u}}_*$, which directly tests the existence argument in Section~5.1 of~\cite{Ven}. There, Chandrasekaran prepares a state in $\mathcal{H}_{\mathcal{D}}$ by projecting the incoming Gaussian $|\Omega_{\mathcal{A}}\rangle$ onto readings below $\ell(\mathcal{D})$,
\begin{equation}
  |\psi\rangle = \frac{P_{\mathcal{D}}|\Omega_{\mathcal{A}}\rangle}{\|P_{\mathcal{D}}|\Omega_{\mathcal{A}}\rangle\|},
  \label{eq:normalized_state}
\end{equation}
which requires a nonzero denominator (Eq.~(5.20)). Because $\hat{\bar{u}}_*$ contains $e^{\hat{\Xi}_y(v)}$, its vacuum expectation value can diverge even if most readings are short. To handle a heavy tail, Section~5.1 introduces a focusing cutoff $\Xi_* > 0$, replaces $\hat{\Xi}_y(v)$ by $\min(\hat{\Xi}_y(v), \Xi_*)$ to define a bounded-focusing clock $\hat{\bar{u}}_*(\Xi_*)$, and forms the resolvent state with Lagrange multiplier $1/L > 0$ (Eqs.~(5.6) and (C.13)),
\begin{equation}
  |\Psi_L\rangle = \lim_{\Xi_* \to \infty} \frac{L}{L + \hat{\bar{u}}_*(\Xi_*)}|\Omega_{\mathcal{A}}\rangle
  = L(L + \hat{\bar{u}}_*)^{-1} P_{\mathcal{M}}|\Omega_{\mathcal{A}}\rangle,
  \label{eq:resolvent}
\end{equation}
where $P_{\mathcal{M}}$ projects onto $\mathcal{M}$, the closure of the domain of the clock's quadratic form~\eqref{eq:clock_form}. If $1 - \langle \Omega_{\mathcal{A}} | \Psi_L \rangle \to 0$ as $L \to \infty$, then Eqs.~(5.17)--(5.18) imply $\|P_{\mathcal{D}}|\Omega_{\mathcal{A}}\rangle\|^2 \to 1$ as $\ell(\mathcal{D}) \to \infty$ (Eq.~(5.19)).

The continuum limit, however, holds the maximum reading $\ell(\mathcal{D})$ fixed while radial resolution is refined. A uniform lower bound on the spectrum of the regulated clock controls the entire projection onto $[0, \ell(\mathcal{D})]$.

\begin{corollary}[Vanishing bounded-reading sector]\label{cor:clock}
At fixed positive area on the spherical linearized background, retain a rotation-invariant angular cutoff containing radiative modes. Let $R$ index positive radial regularizations of~\eqref{eq:quantum_xi} on a common sector. Assume the regulated fields satisfy~\eqref{eq:bessel} and their scalar commutators on each fixed pair of smooth interior tests converge to the linearized bracket uniformly in $y$, with the same radial regularization used for all endpoint integrals containing the tests. Then for every finite $\ell$, the regulated spectral projection $P_{\ell, R} = \mathbf{1}_{[0,\ell]}(\hat{\bar{u}}_{*,R})$ is eventually zero. Its norm limit, and hence its strong limit, is zero.
\end{corollary}

\begin{proof}
Fix an interior interval below $v_0 < 1$, and choose $N$ and its radial test pairs as in~\eqref{eq:regulated_floor}. For all sufficiently large $R$, the sum of the absolute regulated pair commutators is at least $\Lambda_N/2$, uniformly in $y$. Inequalities~\eqref{eq:bessel} and~\eqref{eq:regulated_floor} therefore give $\hat{\Xi}_{y,\mathrm{reg}}(v) \ge b\,\mathbf{1}$ for every $v \in [v_0, 1]$, where the scalar $b > 0$ can be made arbitrarily large by increasing $N$, and is uniform in $y$ by spherical symmetry. At finite resolution, this implies $e^{\hat{\Xi}_{y,\mathrm{reg}}(v)} \ge e^b\mathbf{1}$. Restricting the positive integral~\eqref{eq:clock_form} to $[v_0, 1]$ and using $\int_S \hat{\varrho}_S(y)^2 \epsilon_0 = \mathcal{A}\mathbf{1}$ together with $\hat{a}^2 \le 4\pi\mathcal{A}\mathbf{1}$ (as in Eqs.~(5.21a)--(5.21c)), we obtain
\[
  \hat{\bar{u}}_{*,\mathrm{reg}}
  \ge (1 - v_0)e^b \mathcal{A}\hat{a}^{-1}
  \ge (1 - v_0)e^b \sqrt{\mathcal{A}/(4\pi)}\,\mathbf{1}.
\]
For any prescribed finite $\ell$, the lower edge of the spectrum eventually exceeds $\ell$, forcing $P_{\ell, R} = 0$. The same conclusion already follows from $e^x \ge 1 + x$: replacing $e^b$ by $1 + b$ still sends the lower edge to infinity.
\end{proof}

The same scalar floor sends $L(L + \hat{\bar{u}}_{*,R})^{-1}$ to zero in norm for each fixed $L > 0$, directly identifying the vanishing of the finite-reading sector in the radial limit. In the extended-form convention of Appendix~C.2 of~\cite{Ven}, this gives $\mathcal{M} = \{0\}$ on the linearized sector; by Eq.~(C.13), the projections onto finite readings extend by zero to $\mathcal{M}^\perp$, so $|\Psi_L\rangle = 0$ for every $L > 0$. Consequently, the probability in the linearized counterpart of Eq.~(5.19) vanishes for every finite $\ell(\mathcal{D})$, and the dense form domain inferred at the end of Appendix~C.2 collapses to $\{0\}$.

This discrepancy arises because the fixed-resolution estimates in Appendix~C.2 supply no bound uniform in radial resolution. Whether the coupled non-linear clock exhibits the same collapse remains open. Specifically, the finiteness estimate in Eq.~(5.8) is made ``at fixed angular and radial resolution,'' Eq.~(C.15a) defines the Wigner measure for a ``finite set $\mathcal{L}$ of reduced canonical pairs,'' and the threshold on $L$ in Eq.~(C.16d) depends on the maximum focusing over a compact set at that fixed resolution. Likewise, Eq.~(C.15e) shows only that each coefficient of the Weyl expansion is $O(L^{-1})$ at fixed resolution, without a remainder estimate at fixed $\hbar$ or control uniform in the number of modes. Section~5.1 describes these estimates as ``strong evidence'' for convergence of the resolvent state, and Footnote~18 notes that using the classical tail to bound the exact quantum matrix element is heuristic. Convergence as $L \to \infty$ at each fixed resolution is fully compatible with the opposite behavior when the resolution is removed first. For instance, $N/(L+N) \to 0$ as $L \to \infty$ at fixed $N$, whereas $\lim_{L\to\infty}\lim_{N\to\infty} N/(L+N) = 1$. Corollary~\ref{cor:clock} shows that the positive linearized clock exhibits the latter behavior.

A direct check using the Gaussian Wigner measure of Eq.~(C.15a) confirms the same collapse at the level of phase-space sampling. At fixed angular resolution on the round background, let $Q_n$ denote the classical linearized focusing from the first polarization on a fixed interior interval, truncated to $n$ radial modes. In the centered Gaussian measure of Eq.~(C.15a), diagonalizing the covariance-weighted quadratic form gives $Q_n = \sum_i \mu_{ni} g_i^2$ in law, where the $g_i$ are independent standard normal variables and $\mu_{ni} \ge 0$. Because the Weyl symbol of the square of a linear field is its ordinary square, the mean $\mathbb{E}Q_n$ equals the vacuum expectation value of the regulated positive square. For mode cutoffs resolving the radial tests of Theorem~\ref{thm:linear}, the packet lower bound therefore forces $\mathbb{E}Q_n \to \infty$. For any $t > 0$,
\[
  \mathbb{E}e^{-tQ_n} = \prod_i (1 + 2t\mu_{ni})^{-1/2}
  \le (1 + 2t\,\mathbb{E}Q_n)^{-1/2} \longrightarrow 0,
\]
so $\Pr(Q_n \le M) \le e^{tM}\mathbb{E}e^{-tQ_n} \to 0$ for every finite $M$. This estimate uses only finite-dimensional Gaussian measures, and implies that any almost-sure limit of $Q_n$ on a common probability space is infinite almost surely. The smoothness claim following Eq.~(C.15a) therefore holds only for a finite set of modes; passing to smooth continuum data requires controlling this limit.

\section{What a non-linear quantization would have to cancel}\label{sec:non-linear}

The linearized obstruction of Section~\ref{sec:weyl} relies on two ingredients that behave differently in the full coupled theory. First, the positive ordering in Eq.~(4.22) of~\cite{Ven} still bounds the first polarization's shear square from below even when the radiation and corner data fluctuate together. Second, the shear commutator can acquire field-dependent corrections from parallel transport in field space and from coupling to the corner area. Because positivity survives, we can test directly whether classical non-linear transport removes the short-distance singularity, and then identify what any exact non-linear quantization would have to accomplish in order to admit even a single state of finite focusing.

At fixed local area density, the radiative bracket of Eq.~(A.36) of~\cite{Ven} parallel transports the polarization frame along the null generator in the hyperbolic field space $\mathbb{H}^2$ with metric $\mathsf{g} = \operatorname{diag}(1, e^{2z^1})$. In the orthonormal coframe $e^1 = dz^1$, $e^2 = e^{z^1}dz^2$, parallel transport from $w'$ to $w$ rotates the polarization frame by the angle $\Phi(w,w') = \int_{w'}^w m(t)\,dt$, driven solely by the shear $m(w) \equiv e^{z^1}\partial_w z^2$ of the second polarization. On an angular lattice of cell solid angle $\Delta\Omega$ with effective Planck coefficient $\gamma_y^{\rm lat} = 4\pi\ell_P^2/(\varrho_S(y)^2\Delta\Omega)$, the $z^1$--$z^1$ bracket takes the closed form (derived in Appendix~\ref{sec:geometry})
\begin{equation}
  \{z^1(w), z^1(w')\}
  = -\frac{\gamma_y^{\rm lat}}{\hbar}\,\frac{\operatorname{sgn}(w-w')\cos\Phi(w,w')}{w w'}
  + k_\partial^{11}(w,w').
  \label{eq:z1z1}
\end{equation}
At coincidence ($w = w'$), the rotation angle vanishes, $\Phi(w, w) = 0$, so $\cos\Phi(w,w) = 1$ and the first derivatives of $\cos\Phi$ vanish there as well. Differentiating~\eqref{eq:z1z1} in both arguments therefore leaves the leading linearized $\delta'(w-w')$ coefficient completely unchanged. Appendix~\ref{sec:geometry} isolates the remaining four-term transport kernel $T$ and bounds it in Hilbert--Schmidt norm by $\|T\|_{\rm HS} \le C_\epsilon \Xi_y^{(2)}$, where $\Xi_y^{(2)}$ is the classical focusing due to the second polarization. In particular, $|\langle F_h, T G_h \rangle| \le \|T\|_{\rm HS}$ on every finite-focusing classical configuration. Multiplying by $h$ eliminates this bounded correction as $h \to 0$, leaving the nonzero linearized packet coefficient intact. Classical parallel transport at fixed local area density cannot cancel the short-distance singularity.

Turning from classical transport to the quantum theory, the positivity of the focusing form yields a sharp local condition that any non-linear quantization must satisfy.

\begin{corollary}[Local test for finite focusing]\label{cor:local}
Let $F_h, G_h$ be the rescaled unit-norm profiles of~\eqref{eq:packet} shrinking to an interior point of a resolved generator. On every normalized vector $\psi$ of finite unsubtracted focusing,
\[
  \lim_{h\downarrow 0} \langle [\mathsf{S}(F_h), \mathsf{S}(G_h)]/i \rangle_\psi = 0.
\]
Consequently, to rule out finite focusing on a candidate normalized vector, it suffices to find one pair and a sequence of widths $h \downarrow 0$ along which the absolute commutator expectation stays bounded away from zero, with each fixed smooth test resolved before taking $h \downarrow 0$.
\end{corollary}

\begin{proof}
Define the linear map $T_\psi f = \mathsf{S}(f)\psi$ on real radial test functions supported in $(\epsilon, v)$. The single-test bound in~\eqref{eq:bessel} extends $T_\psi$ continuously to real $L^2(\epsilon, v)$. Applying~\eqref{eq:bessel} to finite subsets of a smooth orthonormal basis $\{e_j\}$ gives $\sum_j \|T_\psi e_j\|^2 \le 4q_\psi < \infty$, so $T_\psi$ is Hilbert--Schmidt and therefore compact. Let $I_h$ be an interval containing the supports of $F_h$ and $G_h$ that shrinks to their center as $h \downarrow 0$. For every $g \in L^2(\epsilon, v)$, Cauchy--Schwarz gives $|\langle g, F_h \rangle| \le \|g\|_{L^2(I_h)} \to 0$, and likewise for $G_h$; hence $F_h$ and $G_h$ converge weakly to zero. Compactness of $T_\psi$ then implies $\|T_\psi F_h\|, \|T_\psi G_h\| \to 0$. Because $|\langle [\mathsf{S}(F_h), \mathsf{S}(G_h)] \rangle_\psi| \le 2\|T_\psi F_h\|\|T_\psi G_h\|$, the commutator expectation vanishes as $h \downarrow 0$.
\end{proof}

The full clock form~\eqref{eq:clock_form} yields a quantitative, background-independent version of this condition. Fix $v_0 \in (\epsilon, 1)$, let $R = \sqrt{\mathcal{A}/(4\pi)}$ be the areal radius of the corner, and let $\psi$ be a unit vector accepted at clock reading $\ell$ in a fixed positive-area sector. Set $\phi_y = \hat{\varrho}_S(y)\hat{a}^{-1/2}\psi$. For any finite orthonormal family of radial test pairs supported below $v_0$, let $C_{k,y}(\phi_y)$ denote the commutator form of the $k$th pair divided by $i$, evaluated on the unnormalized vector $\phi_y$. Then
\begin{equation}
  \int_S \epsilon_0(y) \sum_k |C_{k,y}(\phi_y)|
  \le \frac{4}{1 - v_0}\bigl[\ell - (1 - \epsilon)R\bigr].
  \label{eq:accepted-packets}
\end{equation}
To see this, note that $e^x \ge 1 + x$ for $x \ge 0$. In~\eqref{eq:clock_form}, the constant term contributes $(1 - \epsilon)\mathcal{A}\|\hat{a}^{-1/2}\psi\|^2 \ge (1 - \epsilon)R$ by Cauchy--Schwarz on $S$. For every endpoint $v \ge v_0$, inequality~\eqref{eq:bessel} and the uncertainty relation bound $\sum_k |C_{k,y}(\phi_y)|$ by four times the focusing form at $v$ on $\phi_y$; integrating over $v \in [v_0, 1]$ and $y \in S$ yields~\eqref{eq:accepted-packets}. This bound uses only the ordered clock form on its domain and requires no expansion about a round background. Any continuum realization of the positive, unsubtracted clock that admits a state below reading $\ell$ must therefore explain how its short-distance shear commutator satisfies~\eqref{eq:accepted-packets}.

Where can a coupled quantum realization still depart from~\eqref{eq:z1z1}? Two effects enter beyond fixed-density classical transport: fluctuations of the local area density and quantum operator ordering. On the classical side, a derivative-counting check indicates that area-density fluctuations are subleading in the interior. On a smooth interior background, insert an oscillatory source of radial wave number $k$ into the coupled inversion equations, Eq.~(A.41) of~\cite{Ven}. Counting radial derivatives in the particular solution, the direct radiative response scales as $k^{-1}$ times the source, whereas the induced area variation carries two inverse derivatives. Feeding that area variation back into the first-order transport equation contributes at order $k^{-3}$, as does the auxiliary variable $b$ when reconstructing the full radiative variation $U$. Although this interior derivative counting explains why area feedback is expected to be subleading, homogeneous matching solutions at the cuts and controlling coincident products of operators at fixed $\hbar$ remain open in the full coupled sector.

To make explicit which quantum estimate would settle the coupled sector, keep the angular regulator fixed and take the $N/2$ radial test pairs $(X_k, Y_k)$ of Section~\ref{sec:weyl} at fixed $y$ and $v$, with linearized commutators $i\lambda_k$ satisfying~\eqref{eq:regulated_floor}. Fix a scalar reference coefficient $\gamma_y > 0$ (absorbing any fluctuations of the coefficient into the remainder), and for a normalized vector $\psi$ define the real remainder $r_k(\psi)$ via commutator forms by $\langle [X_k, Y_k]/i \rangle_\psi = \lambda_k + r_k(\psi)$.

\begin{assumption}[Ordering on radial tests]\label{ass:O}
For each normalized vector $\psi$ with a finite candidate focusing bound $q_\psi$, the remainder of the full smeared commutator obeys
\begin{equation}
  \sum_{k \le N/2} |r_k(\psi)|
  \le \gamma_y\big[(c_2\sqrt{N} + c_3)q_\psi + B_\psi\big],
  \qquad c_2 = C_\epsilon/\sqrt{2},
  \label{eq:O}
\end{equation}
where $C_\epsilon = \epsilon^{-2}(2 + \sqrt{2\log(1/\epsilon)})^2$ is the transport constant of Appendix~\ref{sec:geometry}, and $c_3$ and $B_\psi < \infty$ are independent of $N$. In a regulated realization, this estimate is required for the resolved tests uniformly as radial resolution is removed along any sequence on which the focusing of $\psi$ remains bounded by $q_\psi$. The remainder includes ordering, boundary, and edge-mode contributions.
\end{assumption}

\begin{theorem}[Conditional non-linear no-go]\label{thm:non-linear}
For the positive, unsubtracted focusing form, suppose the reference brackets satisfy~\eqref{eq:regulated_floor} and the full remainder satisfies Assumption~\ref{ass:O}. Then the finite-focusing domain is trivial. The same conclusion holds under the weaker hypothesis that $\sum_{k \le N/2} |r_k(\psi)| = o(\Lambda_N)$ on each candidate finite-focusing vector $\psi$.
\end{theorem}

The proof of Theorem~\ref{thm:non-linear}, together with the necessary cancellation bound (Proposition~\ref{prop:cancel}) and the extension to the corner-weighted clock~\eqref{eq:clock_form}, is given in Appendix~\ref{app:many}. Although the classical transport bound~\eqref{eq:HS} of Appendix~\ref{sec:geometry} motivates~\eqref{eq:O} at leading quadratic order, verifying this estimate for quantum operator products at fixed $\hbar$ with matched, fluctuating corner data remains the open step.

\section{Gluing, closed norms, and the dimension count}\label{sec:hidden}

We now turn to the second stage of Chandrasekaran's argument. For this section, we grant that the exact clock admits a nonzero accepted space $\mathcal{H}_{\mathcal{D}}$ and enforces the corner-area ceiling~\eqref{eq:area_bound}, $0 \le \hat{s} \le s_*$. Section~5.2 of~\cite{Ven} seeks to prove $\dim \mathcal{H}_{\mathcal{D}} < \infty$ by comparing a cut preparation with the same preparation on the uncut diamond.

Once the corner area is bounded, each intermediate state contributes a weight bounded away from zero to the closed sum. As a result, finiteness of the squared norm of the closed uncut preparation is already equivalent to finite dimensionality, with no boost argument needed (Proposition~\ref{prop:trace}). Passing to normalized probabilities does not evade this requirement. In infinite dimension, normalized cut probabilities vanish on every fixed finite-rank measurement and therefore cannot reproduce any normal state on the uncut geometry (Corollary~\ref{cor:normal-gluing}).

To see how this enters Section~5.2 of~\cite{Ven}, recall that the operator $C_b$ prepares the future half of the causal diamond, weighted by a positive Gaussian factor $f_b(\hat{s})$ centered on the dimensionless area of a smooth reference geometry (Eqs.~(5.23)--(5.25)). Gluing $C_b$ to its time reflection along the future cap yields the positive operator $C_b^\dagger C_b$ on $\mathcal{H}_{\mathcal{D}}$, and closing the remaining cut at the corner $S$ sums its diagonal matrix elements. Because $0 \le \hat{s} \le s_*$ on $\mathcal{H}_{\mathcal{D}}$, the minimum $m_{b,\mathcal{D}} = \min_{0 \le s \le s_*} f_b(s)$ is strictly positive, giving the uniform operator bound of Eq.~(5.30c):
\[
  c\,\mathbf{1} \le C_b^\dagger C_b \le \mathbf{1}, \qquad c \equiv e^{-s_*} m_{b,\mathcal{D}} > 0.
\]

Section~5.2 argues by contradiction. Assume $\dim \mathcal{H}_{\mathcal{D}} = \infty$, and partition the compact interval $[0, s_*]$ into finitely many narrow bands of width $\Delta s$. At least one band $I \subset [0, s_*]$ must contain infinitely many orthonormal states. On states supported in $I$, a relative corner boost of rapidity $\eta$, generated by $\hat{\mathcal{K}} = \hat{\mathcal{A}}/(8\pi G_N)$ via $e^{i\eta \hat{s}/(2\pi)}$, acts by nearly a constant phase and therefore cannot reproduce the boost dependence of a two-sided graviton correlation function across $S$ (Eq.~(5.29)). Grouping all basis states of $\mathcal{H}_{\mathcal{D}}$ into finite cycles dominated by states in $I$, and letting $P$ project onto a finite union of such cycles, yields unnormalized partial-sum amplitudes $Z_P[X] \le 0$ for a bounded two-sided observable $X$, whereas the uncut preparation gives $Z[X] > 0$ (Eq.~(5.38)).

The sign difference $Z_P[X] \le 0 < Z[X]$ produces a contradiction only if the cut sums converge to the finite uncut amplitude. Before inserting the signed observable $X$, however, one can close the same cut with the identity inserted. Every term then has a definite sign, and the uniform lower bound on $C_b^\dagger C_b$ immediately ties the closed norm to the dimension of $\mathcal{H}_{\mathcal{D}}$.

\begin{proposition}[Closed norm on the bounded-area sector]\label{prop:trace}
Under the uniform weight bound $c\,\mathbf{1} \le C_b^\dagger C_b \le \mathbf{1}$ of Eq.~(5.30c) of~\cite{Ven}, the preparation $C_b$ is Hilbert--Schmidt; that is, $\operatorname{Tr}(C_b^\dagger C_b) < \infty$, if and only if $\mathcal{H}_{\mathcal{D}}$ is finite-dimensional.
\end{proposition}

\begin{proof}
For any $n$ orthonormal vectors in $\mathcal{H}_{\mathcal{D}}$, the sum of the diagonal matrix elements of $C_b^\dagger C_b$ lies between $cn$ and $n$. Taking the supremum over all finite orthonormal sets gives
\[
  c\,\dim \mathcal{H}_{\mathcal{D}}
  \le \operatorname{Tr}(C_b^\dagger C_b)
  \le \dim \mathcal{H}_{\mathcal{D}}.
\]
\end{proof}

If identity gluing computes a finite squared norm for the uncut preparation, Proposition~\ref{prop:trace} already implies $\dim \mathcal{H}_{\mathcal{D}} < \infty$ without invoking any boost. Equation~(5.66) of~\cite{Ven} expresses the thermal partition function as an integral of $\operatorname{Tr}(C_b^\dagger C_b)$ over the reference area parameter; justifying the finiteness of that trace is therefore equivalent to the counting problem itself. Importantly, the convergence bound in Eq.~(4.30) of~\cite{Ven} does not supply this finiteness: Eq.~(4.30) uses Cauchy--Schwarz to control the composition of amplitudes between specified normalized boundary vectors, whereas closing a preparation sums its diagonal weights. Even the identity operator on $\ell^2$ has bounded matrix elements between unit vectors while having infinite trace.

A simple example illustrates why an area ceiling and a boost phase alone cannot rule out infinite dimension. At fixed corner area, consider the quadratic Fock space and a modified clock that retains only angular harmonics $j \le J$ of the shear before squaring it. If the corner factors also act only on the retained sector, the clock projection on $\mathcal{K}_{\le J} \otimes \mathcal{F}_{>J}$ factorizes as $P_{\le J} \otimes \mathbf{1}$. Any accepted low-mode state tensored with orthogonal high-mode excitations then gives infinitely many orthonormal states with identical area, clock probabilities, and boost phase. Under the weights of Eq.~(5.30c), each state contributes at least $c > 0$ to the identity closure, producing a divergent trace, which is the same divergence that Chandrasekaran exhibits for matter states in JT gravity in Eq.~(2.17b). For this clock, the area ceiling and the cycle construction coexist with $\dim \mathcal{H}_{\mathcal{D}} = \infty$. Applying Eq.~(5.38) still requires independently identifying the cut limit with a finite uncut amplitude.

One might ask whether normalizing the cut preparation allows the cut probabilities to converge even when the unnormalized trace diverges. To test this, set $W = C_b^\dagger C_b$ on the nonzero accepted space $\mathcal{H}_{\mathcal{D}}$, and consider a fixed finite-rank projection $Q$ measured on the conjugate copy of the cut. Identifying that copy with the conjugate of $\mathcal{H}_{\mathcal{D}}$, Eq.~(D.22) of~\cite{Ven} assigns $Q$ the normalized cut probability $\operatorname{Tr}(\rho_P Q)$, where $\rho_P = PWP/\operatorname{Tr}(PWP)$ is the density matrix of Eq.~(5.32).

\begin{corollary}[Normalized gluing and finite-rank measurements]\label{cor:normal-gluing}
Let $W = C_b^\dagger C_b$ on the nonzero accepted space $\mathcal{H}_{\mathcal{D}}$ and assume $c\,\mathbf{1} \le W \le \mathbf{1}$ with $c > 0$ as in Eq.~(5.30c) of~\cite{Ven}. Let $P$ run over an increasing family of finite-rank projections tending strongly to the identity, and set $\rho_P = PWP/\operatorname{Tr}(PWP)$. If $\mathcal{H}_{\mathcal{D}}$ is infinite-dimensional, then $\operatorname{Tr}(\rho_P Q) \to 0$ for every fixed finite-rank projection $Q$. In particular, agreement of the limiting cut probabilities with any normal uncut marginal on $\mathcal{B}(\mathcal{H}_{\mathcal{D}})$ is possible only in finite dimension.

More quantitatively, suppose normalized gluing holds for the single rank-one test $Q = |\chi\rangle\langle\chi|$ of a specified accepted unit vector $\chi \in \mathcal{H}_{\mathcal{D}}$, and that its uncut probability is $p_\chi > 0$. Writing $w_\chi = \langle \chi | W | \chi \rangle$, the dimension $d = \dim \mathcal{H}_{\mathcal{D}}$ satisfies
\[
  \frac{w_\chi}{p_\chi} \le d \le \frac{w_\chi}{c\,p_\chi}.
\]
\end{corollary}

\begin{proof}
For $n = \operatorname{rank} P > 0$, we have $PWP \le P$ and $\operatorname{Tr}(PWP) \ge cn$, so any rank-$r$ projection $Q$ satisfies $\operatorname{Tr}(\rho_P Q) \le r/(cn)$. In infinite dimension $n \to \infty$, so $\operatorname{Tr}(\rho_P Q) \to 0$. Because any normal state on $\mathcal{B}(\mathcal{H}_{\mathcal{D}})$ is represented by a density matrix, it must assign positive expectation to some rank-one projection; a single positive limiting probability therefore excludes infinite dimension. In finite dimension, the increasing projections $P$ eventually equal the identity, giving $p_\chi = w_\chi / \operatorname{Tr} W$. Combining this with $cd \le \operatorname{Tr} W \le d$ yields the stated bounds.
\end{proof}

Existence of an accepted unit vector $\chi$ already provides an open-boundary preparation with $w_\chi > 0$. Independently establishing $p_\chi > 0$ in the uncut geometry and showing that it equals the normalized cut limit would therefore yield a direct state count, independent of the choice of cycles in Eq.~(5.38).

The distinction between defining a closure and establishing its finiteness also appears in Appendix~E of~\cite{Ven}. Condition~5 requires finite amplitudes for compatible closures of allowed smeared preparations, and Eq.~(E.15b) identifies the squared norm of $C_b$ with $\operatorname{Tr}(C_b^\dagger C_b)$. However, the verification of Condition~5 for the causal diamond in Eq.~(E.17) invokes the finite-dimension result from Section~5.2, so Appendix~E does not supply an independent proof of finiteness.

A potential source of independent input is the one-loop calculation in Appendix~D.3 of~\cite{Ven}. By Eq.~(D.47), if the ratio of the cyclic two-copy trace to the square of the one-copy trace has a positive limit as all intermediate states are included, that limit bounds the one-copy trace itself. Equations~(D.48)--(D.49) obtain a positive lower bound for this ratio on the regular family of saddles, using the occupation-number count of Eq.~(D.45) with cap data, corner data, and transverse expansion specified so that the stronger focusing form in Eq.~(D.40) is bounded. Extending that lower bound to the full trace selected by the incoming clock would supply the independent finiteness input that the gluing proof requires.

\section{Infinite rank of the vacuum-subtracted quadratic clock}\label{sec:subtracted-clock}

The continuum obstruction in Section~\ref{sec:weyl} arises because the positive focusing operator retains the zero-point energy of the shear. A natural repair is to subtract the state-independent vacuum contribution from the first non-vanishing radiative correction to the affine time. This defines a different spectral problem. Even if subtraction yields a lower-bounded quantum clock, bounding its reading can select finitely many states only if gravitational perturbations of arbitrarily small delay are sufficiently restricted.

In this section, we show that, for any self-adjoint realization of the vacuum-subtracted quadratic clock obeying standard canonical displacement identities, every spectral threshold strictly above the minimum reading admits infinitely many orthogonal states (see Theorem~\ref{thm:subtracted-clock}). We then show why this obstruction is specific to the quadratic approximation. The same localized perturbations produce divergent exact classical readings at fixed amplitude, while rescaling their amplitudes to keep the exact readings bounded drives their symplectic pairing to zero.

The mechanism behind Theorem~\ref{thm:subtracted-clock} is a mismatch between how the quadratic delay and the symplectic form weight radiation near the corner $v = 1$. On the round background of fixed positive area, a shear perturbation at areal radius $v$ delays the null generator only over the remaining interval of length $1 - v$. Consequently, the quadratic affine delay carries the radial weight $v(1 - v)$, which vanishes at the corner, whereas the radial weight $v^2$ in the radiative symplectic pairing $\Omega$ remains strictly positive there~\eqref{eq:delay-pairing}. This mismatch allows one to construct pairs of radiative perturbations in a single angular harmonic whose quadratic delays tend to zero while their mutual symplectic pairing $\Omega$ approaches a nonzero constant. Appendix~\ref{app:packets} constructs such pairs~\eqref{eq:soft-packets} subject to the linearized endpoint and matching conditions.

To translate these classical perturbations into a statement about the quantum clock, consider the response of a vacuum-subtracted quadratic clock to a canonical displacement by a profile and its opposite. Because any state-independent subtraction cancels identically from the second difference between opposite displacements, the net shift equals twice the classical quadratic delay~\eqref{eq:displacement-delay}. If this displacement identity extends to the form domain of the quantum clock, the two localized perturbations retain their canonical commutation relation arbitrarily close to the bottom of the spectrum.

\begin{theorem}[Infinite rank above the minimum quadratic reading]\label{thm:subtracted-clock}
Let $U$ be a lower-bounded self-adjoint realization of the quadratic clock, with spectrum $\sigma(U)$. Suppose the localized displacements in~\eqref{eq:soft-packets} and their real multiples act on its Hilbert space by Weyl unitaries with symplectic pairing $\Omega$, preserve the form domain of $U$, and obey the displacement identity~\eqref{eq:displacement-delay} there. Then for every threshold $\ell > \inf\sigma(U)$, the sublevel projection $\mathbf{1}_{(-\infty,\ell]}(U)$ has infinite rank.
\end{theorem}

The argument is by contradiction, with full details given in Appendix~\ref{app:packets}. If $\mathbf{1}_{(-\infty,\ell]}(U)$ had finite rank for some $\ell > \inf\sigma(U)$, then $\inf\sigma(U)$ would be an eigenvalue with a finite-dimensional ground space separated by a positive spectral gap from its orthogonal complement. By the displacement identity~\eqref{eq:displacement-delay}, acting on a ground state with either localized packet costs a vanishingly small increase in clock reading, so the spectral gap forces the displaced states back into the ground space in the limit. Compressing the Weyl unitaries to this finite-dimensional ground space therefore yields two limiting unitaries obeying a Weyl commutation relation. Rescaling one packet adjusts the limiting Weyl phase to an irrational multiple of $2\pi$, which contradicts taking determinants on a finite-dimensional space.

Applying Theorem~\ref{thm:subtracted-clock} to a physical clock requires both establishing lower boundedness and controlling the sign of the readings. For the actual quadratic delay weight $w(v) = v(1 - v)$, the derivative term $|\partial_v \sqrt{w}|^2$ behaves asymptotically as $1/[4(1 - v)]$ near the corner $v = 1$. Its integral diverges logarithmically, leaving lower boundedness open under the smooth-sampling quantum energy inequality discussed in Appendix~\ref{sec:repairs}. Furthermore, if $U$ is nonnegative, the sublevel $(-\infty, \ell]$ in Theorem~\ref{thm:subtracted-clock} coincides with the physical interval $[0, \ell]$. If vacuum subtraction permits negative readings, however, the infinite rank could reside entirely in the negative spectrum, so bounding the rank on $[0, \ell]$ would require separate control. The theorem also leaves open the multiplicity at $\inf\sigma(U)$ itself.

Finally, passing from the quadratic delay to the exact non-linear affine time reveals a sharp limitation of these packets. Concentrating a perturbation near the corner, controlled by the localization parameter $L \to \infty$ in Appendix~\ref{app:packets}, makes the remaining areal interval $1 - v$ exponentially short, but the exact clock integrand~\eqref{eq:classical_clock} contains the exponential $e^{\Xi_y(v)}$ of the accumulated focusing. On the incoming profiles of~\eqref{eq:soft-packets} at any fixed nonzero amplitude, the accumulated focusing grows at least as $e^{2L}/L$, overwhelming the shrinking integration interval and causing the exact classical reading to diverge. Conversely, shrinking both amplitudes with $L$ to keep the exact readings below a fixed threshold sends their symplectic pairing $\Omega$ to zero~\eqref{eq:scaled-pairing}, eliminating the nontrivial Weyl phase on which Theorem~\ref{thm:subtracted-clock} relies. Because this comparison evaluates exact incoming profiles with matching to the opposite sheet checked only to first order, the rank of a renormalized version of the exact non-linear clock remains open.

This naturally raises the question of whether one can instead normal-order or smear $\hat{\Xi}_y(v)$ directly inside the exponential of~\eqref{eq:clock_form}, because strict positivity is not required. A uniform lower bound $\hat{\Xi}_y(v) \ge -C\mathbf{1}$ would still preserve the area ceiling~\eqref{eq:area_bound} up to a factor $e^{2C}$. Appendix~\ref{sec:repairs} shows why standard quadratic-field prescriptions fail to achieve this, either diverging to $-\infty$ across angular harmonics when smeared after squaring or leaving infinitely many high-$j$ modes unconstrained when smeared before squaring.

\section{Discussion}\label{sec:discussion}

Taken together, our results show that~\cite{Ven} does not establish the existence of a nonzero, finite-dimensional continuum Hilbert space for a causal diamond. Instead, the gravitational clock is caught in a structural tension. Retaining the unsubtracted positive shear square preserves the classical area ceiling, but on a round background of fixed positive area the linearized null commutator drives the entire regulated clock spectrum past every finite reading as radial resolution is refined (Section~\ref{sec:weyl}). The fixed-resolution existence estimate in Appendix~C.2 of~\cite{Ven} supplies no bound uniform in radial resolution, leaving open whether the coupled non-linear clock exhibits the same collapse. Subtracting the vacuum contribution removes the zero-point divergence, but it sacrifices the positivity that bounds the corner area and, for a nonnegative quadratic clock obeying the displacement identities of Theorem~\ref{thm:subtracted-clock}, allows infinitely many orthogonal states below any threshold strictly above the spectral bottom (see Section~\ref{sec:subtracted-clock}). Because the discrete area spectrum and the finite Lorentzian thermal trace in Section~5 of~\cite{Ven} rest on $\mathcal{H}_{\mathcal{D}}$ being nontrivial and finite-dimensional, those consequences remain unestablished in the continuum limit.

Even if one grants that a non-linear or renormalized version of the exact clock selects a nonempty space of states with bounded corner area, the dimension count in Section~5.2 of~\cite{Ven} does not close on its own. Once the corner area is capped, every intermediate state enters the identity closure with a weight bounded away from zero, making finiteness of the closed norm equivalent to finite dimensionality (see Proposition~\ref{prop:trace}). Applying the boost contradiction of Eq.~(5.38) therefore requires an independent proof that the cut sums converge to a finite uncut amplitude. As Section~\ref{sec:hidden} shows, such an input would settle the count directly, either by establishing a finite uncut norm for the identity insertion, by reproducing a single positive normalized probability for a rank-one measurement on the uncut geometry (see Corollary~\ref{cor:normal-gluing}), or by extending the one-loop cyclic-trace bound of Appendix~D.3 to the full clock-selected trace. On the subtraction side, an equal mixture of $N$ orthogonal states below a threshold in Theorem~\ref{thm:subtracted-clock} has entropy $\log N$; turning that into a tension with the covariant entropy bound~\cite{Bousso} would require showing that such mixtures can be prepared on the same exact light-sheet with their non-linear gravitational response included, where our exact classical calculation shows the same packets diverge.

For the positive, unsubtracted clock of Eq.~(4.22) of~\cite{Ven}, the central open question is whether non-linearities and fluctuating corner data can cancel the short-distance shear commutator at fixed $\hbar$. On a resolved generator of nondegenerate area density, we conjecture that they cannot, and that the positive unsubtracted focusing form of Eq.~(4.22) has trivial domain in any regular continuum quantization of Chandrasekaran's full coupled incoming brackets. Corollary~\ref{cor:local} reduces this conjecture to a sharp local test. Finite focusing forces the unscaled commutator expectation on shrinking radial packets to tend to zero, so a nonzero residual along a single sequence of widths on each candidate vector would prove the conjecture, whereas exhibiting a single nonzero vector of finite focusing in the coupled representation would disprove it. Equations~(A.36) and~(A.37)--(A.42) of~\cite{Ven} fix the classical brackets at fixed and varying area density, and Eq.~(D.33) uses the perturbative two-point function of Appendix~D; controlling the exact quantum commutator on matched cuts at fixed $\hbar$ as radial resolution is removed is the decisive test of whether the unsubtracted gravitational clock survives the continuum limit.

\appendix

\section{The non-linear brackets and the ordering estimate}\label{sec:geometry}

This appendix supplies the classical parallel-transport estimates used in Section~\ref{sec:non-linear} and proves the conditional non-linear no-go theorem (Theorem~\ref{thm:non-linear}) by summing over many disjoint pairs. At fixed local area density, the non-linear radiative bracket parallel transports the polarization frame along the null generator. Because the second polarization drives both this frame rotation and the second positive term in the focusing integral, its focusing directly bounds the transport correction to the bracket of the first polarization.

\subsection{Parallel transport at fixed area density}

The angular harmonic cutoff $J$ used in Section~\ref{sec:weyl} is inconvenient in the non-linear theory because products of band-limited fields generate higher harmonics. Following the incoming null generators, we instead introduce a lattice of cells of solid angle $\Delta\Omega$ on $S$. At fixed angular embedding, the radiative part of the symplectic form (Eq.~(A.14) of~\cite{Ven}) contains no angular derivatives, so the transverse delta function reduces to $\delta_{yy'}/\Delta\Omega$. We first hold the local area density fixed and return to the boundary, charge, and embedding terms below.

The two graviton polarizations $z^1, z^2$ are coordinates on the space of unit-determinant conformal metrics on a cut. Equipped with the field-space metric $\mathsf{g} = \operatorname{diag}(1, e^{2z^1})$ of Eq.~(4.21c) of~\cite{Ven}, this space is the hyperbolic plane $\mathbb{H}^2$, with a natural orthonormal coframe $e^1 = dz^1$, $e^2 = e^{z^1}dz^2$. In this frame, the shear of the second polarization is
\begin{equation}
  m(w) \equiv e^{z^1}\partial_w z^2,
\end{equation}
and the $z^2$ contribution to the classical focusing integral~\eqref{eq:classical_xi} is
\begin{equation}
  \Xi^{(2)}_y(v) \equiv \frac{1}{4}\int_\epsilon^v w\,m(w)^2\,dw \le \Xi_y(v).
  \label{eq:Xi2}
\end{equation}
This quantity controls the classical correction from parallel transport.

Write $\tau_v = \log(v/\epsilon)$, and recall the corner charge aspect $\mathcal{K}(y) \equiv \varrho_S(y)^2/(8\pi G_N)$ from~\cite{Ven} and the Planck length $\ell_P^2 = G_N\hbar$. Up to overall normalization, the radiative symplectic form along a generator is (Eqs.~(A.14)--(A.15) of~\cite{Ven})
\begin{equation}
  \varrho_S(y)^2 \int_\epsilon^1 w^2\,\mathsf{g}_{\alpha\beta}\,\nabla_w(\delta^o z)^\alpha \wedge (\delta^o z)^\beta\,dw,
  \qquad
  \delta^o z \equiv \delta z - \frac{w}{2}\,\delta\log\varrho_S^2\,\partial_w z,
\end{equation}
where $\nabla_w$ is the covariant derivative of $\mathsf{g}$ along the curve $w \mapsto z(w)$ and $\delta^o z$ accounts for the shift of the areal coordinate when the corner area changes. Inverting the fixed-area-element radiative block (Eqs.~(A.34)--(A.36) of~\cite{Ven}) on the lattice gives
\begin{equation}
\begin{aligned}
  \{z^\alpha(w,y), z^\beta(w',y')\}
  &= \Big[-\frac{\gamma_y^{\rm lat}}{\hbar}\,\frac{\operatorname{sgn}(w-w')}{w w'}\,
  P^\alpha{}_\mu(w,w')\,\mathsf{g}^{\mu\beta}(z(w'))
  + k_\partial^{\alpha\beta}(w,w')\Big]\delta_{yy'},\\
  \gamma_y^{\rm lat} &\equiv \frac{\hbar}{2\mathcal{K}(y)\Delta\Omega}
  = \frac{4\pi\ell_P^2}{\varrho_S(y)^2\Delta\Omega}.
\end{aligned}
\label{eq:A36}
\end{equation}
Here $P(w,w')$ denotes parallel transport along $z(\cdot)$ in $\mathbb{H}^2$, and $k_\partial$ solves the source-free transport equation in each argument with coefficients fixed by the cut and corner conditions.

The coefficient $\gamma_y^{\rm lat}$ is the ratio of the Planck area to the patch area $\varrho_S(y)^2\Delta\Omega$ represented by generator $y$, and thus acts as an effective $\hbar$ on that generator. Because every patch has area at most the total corner area $\mathcal{A}$, we have the uniform lower bound
\begin{equation}
  \gamma_y^{\rm lat} \ge \frac{4\pi\ell_P^2}{\mathcal{A}} = \frac{\pi}{s}
  \label{eq:gmin}
\end{equation}
on every lattice. We set $\gamma_y = \gamma_y^{\rm lat}$ when applying~\eqref{eq:floor} in a fixed-charge lattice sector.

In the orthonormal frame $(e^1, e^2)$, the connection one-form of $\mathbb{H}^2$ is $\omega^2{}_1 = e^2$, so parallel transport from $w'$ to $w$ rotates the polarization frame by the angle
\begin{equation}
  \Phi(w,w') = \int_{w'}^w m(t)\,dt,
\end{equation}
driven solely by the second polarization's shear $m$. Because the dual frame vector $e_1$ equals $\partial_{z^1}$, evaluating~\eqref{eq:A36} for $\alpha = \beta = 1$ yields the closed form~\eqref{eq:z1z1}. At linear order $\Phi = 0$, recovering the sign kernel of Theorem~\ref{thm:linear}; the fixed-density bulk kernel differs from the linearized one only by the factor $\cos\Phi(w,w')$.

Differentiating~\eqref{eq:z1z1} in both arguments separates the singular part of the classical shear bracket from the parallel-transport correction:
\begin{equation}
  \{\partial_w z^1(w), \partial_{w'}z^1(w')\}
  = -\frac{\gamma_y^{\rm lat}}{\hbar}\big[D_0(w,w') + \operatorname{sgn}(w-w')\,\rho(w,w')\big]
  + \partial_w\partial_{w'}k_\partial^{11},
  \label{eq:JJ}
\end{equation}
where $D_0$ is the linearized bracket ($\Phi \equiv 0$) and
\begin{equation}
  \rho(w,w') = \frac{\cos\Phi - 1}{w^2 w'^2}
- \frac{\sin\Phi\,m(w')}{w^2 w'}
  + \frac{\sin\Phi\,m(w)}{w w'^2}
  + \frac{\cos\Phi\,m(w)m(w')}{w w'}.
  \label{eq:rho}
\end{equation}
At coincidence ($w = w'$), $\Phi = 0$ and the first derivatives of $\cos\Phi$ vanish, so the coefficients of $\delta'(w-w')$ and $\delta(w-w')$ agree identically with their linearized values. At finite classical focusing, $m$ need only be square-integrable on an interior interval, so the natural norm for the remainder kernel is the Hilbert--Schmidt norm. Building the Robertson pairs from $z^1$ alone avoids the field-dependent $\delta(w-w')$ terms that appear in brackets involving $m$ (reflecting the non-abelian isometry group of $\mathbb{H}^2$), while the focusing contribution~\eqref{eq:Xi2} of $m$ directly bounds the rotation $\Phi$.

For each orthonormal pair of real radial profiles $(a_k, b_k)$ used in~\eqref{eq:regulated_floor}, the classical transport contribution $R_k^{\rm cl}$ (in commutator units) is, up to sign, $\gamma_y^{\rm lat}\langle a_k, T b_k \rangle$, where $T(w,w') = \sqrt{ww'}\,\operatorname{sgn}(w-w')\rho(w,w')$. By Cauchy--Schwarz, $|\Phi| \le 2(\tau_v \Xi^{(2)})^{1/2}$, which implies $|\cos\Phi - 1| \le 2\tau_v \Xi^{(2)}$ and $|\sin\Phi| \le 2(\tau_v \Xi^{(2)})^{1/2}$. Setting $a(w) = w^{-3/2}$ and $d(w) = m(w)/\sqrt{w}$ on $[\epsilon, v]$, we have $\|a\|_2 \le 1/(\sqrt{2}\,\epsilon)$ and $\|d\|_2 \le 2\sqrt{\Xi^{(2)}}/\epsilon$. Bounding the four terms of $T$ in $L^2([\epsilon,v]^2)$ with these norms yields
\[
  \|T\|_{\rm HS} \le \epsilon^{-2}\bigl(\tau_v + 4\sqrt{2\tau_v} + 4\bigr)\Xi^{(2)}_y(v)
  \le C_\epsilon\,\Xi^{(2)}_y(v),
  \qquad C_\epsilon \equiv \epsilon^{-2}\bigl(2 + \sqrt{2\log(1/\epsilon)}\bigr)^2,
\]
uniformly for $v \le 1$. Applying Cauchy--Schwarz in $k$ and Bessel's inequality to the orthonormal family $\{a_k, b_k\}$ then gives
\begin{equation}
  \sum_{k \le N/2} |R_k^{\rm cl}|
  \le \gamma_y^{\rm lat}\sqrt{N/2}\,\|T\|_{\rm HS}
  \le \gamma_y^{\rm lat}\sqrt{N/2}\,C_\epsilon\,\Xi^{(2)}_y(v).
  \label{eq:HS}
\end{equation}
At leading quadratic order, a positive semidefinite quadratic form in real canonical variables quantizes to a sum of squares; applying this to the signed sums in~\eqref{eq:HS} motivates~\eqref{eq:O} at that order, though it does not control higher field powers or the full Moyal commutator.

\subsection{Boundary, charge, and embedding terms}

Next, we check how the cut conditions, the corner charge, and the angular embedding affect the sum over $N/2$ orthonormal radial pairs. A solution $k_\partial$ of the source-free transport equation in both arguments factors through the two-dimensional polarization space at a reference cut $w_*$. In the transported orthonormal frame,
\[
  k_\partial(w,t) = (wt)^{-1} P(w, w_*)\,M_\partial\,P(w_*, t),
\]
where the $2\times 2$ matrix $M_\partial$ is fixed by the cut conditions. Its radial rank is therefore at most two, and its differentiated profiles are linear combinations of $\partial_w[w^{-1}\cos\Phi(w, w_*)]$ and $\partial_w[w^{-1}\sin\Phi(w, w_*)]$, which have finite weighted $L^2$ norm on finite-focusing data. More generally, for any finite-rank radial kernel $\sum_{i=1}^r u_i(w)\tilde{u}_i(w')$, Bessel's inequality bounds the sum of the absolute pair contributions by
\[
  \sum_{i=1}^r \|\sqrt{w}\,\partial_w u_i\|_2\,\|\sqrt{w}\,\partial_w \tilde{u}_i\|_2,
\]
independent of the number of radial tests $N$.

At fixed angular embedding, write $\kappa = \varrho_S^2$. In a local chart where the radiative symplectic form takes the shape $\kappa\,\omega_{\rm rad} + \delta\kappa \wedge \beta$, closedness forces $\delta_{\rm rad}\beta = \omega_{\rm rad}$. Passing to radiative Darboux coordinates $(q, p)$, one can absorb the exact part of $\beta$ by shifting the variable conjugate to $\kappa$, leaving canonical radiative momenta $P = \kappa p$. Any radiative observable depending on $q, P/\kappa, \kappa$ is then independent of the conjugate of $\kappa$. Its Poisson bracket with $\kappa$ vanishes, and in a Weyl calculus in which $\kappa$ is a linear coordinate, its Moyal bracket with $\kappa$ vanishes too.

Allowing the angular embedding and local area density to fluctuate introduces further couplings. Even when the incoming labels follow the incoming cone, the corner angular momentum involves the normal connection and the matching condition in Eq.~(A.29) of~\cite{Ven}, while Eqs.~(A.37)--(A.42) govern the coupled inversion when the area element varies. Finite-rank classical control applies to contributions factoring through finitely many corner functionals of the regularity above; in Theorem~\ref{thm:non-linear}, all remaining coupled and operator-ordering terms are absorbed into the commutator remainder $r_k(\psi)$.

\subsection{Many-pair cancellation and proof of Theorem~\ref{thm:non-linear}}\label{app:many}

With the radial test pairs $(X_k, Y_k)$ and commutator remainders $r_k(\psi)$ defined in Section~\ref{sec:non-linear}, we first record the necessary cancellation condition on any finite-focusing state and then prove Theorem~\ref{thm:non-linear}.

\begin{proposition}[Necessary cancellation]\label{prop:cancel}
Every normalized vector $\psi$ of finite unsubtracted focusing $q_\psi$ satisfies
\begin{equation}
  \sum_{k \le N/2} |\lambda_k + r_k(\psi)| \le 4q_\psi,
  \qquad
  \left|\Lambda_N + \sum_{k \le N/2} r_k(\psi)\right| \le 4q_\psi
  \label{eq:cancellation}
\end{equation}
for every $N$.
\end{proposition}

\begin{proof}
Applying~\eqref{eq:robertson} to each pair $(X_k, Y_k)$ with its full commutator and summing over $k \le N/2$ gives the first inequality via~\eqref{eq:bessel}, because all $N$ test functions are orthonormal. The triangle inequality then yields the second bound.
\end{proof}

Thus, on any finite-focusing state, the non-linear remainder must have a signed sum equal to $-\Lambda_N + O(q_\psi)$, and the total absolute residual must remain bounded as $N$ grows. Note that this leaves room for substantial renormalization of the bracket. Even replacing each $\lambda_k$ by $\lambda_k/2$ produces a remainder of order $\Lambda_N$ while still forcing the focusing to diverge. The $\sqrt{N}$ bound in Assumption~\ref{ass:O}, motivated by~\eqref{eq:HS}, is therefore stronger than the minimal condition required to rule out finite focusing.

\begin{proof}[Proof of Theorem~\ref{thm:non-linear}]
By Proposition~\ref{prop:cancel} and Assumption~\ref{ass:O}, any finite-focusing vector $\psi$ must satisfy
\[
  4q_\psi \ge \Lambda_N - \sum_{k \le N/2} |r_k(\psi)|
  \ge \gamma_y c_1 N^2 - \gamma_y\big[(c_2\sqrt{N} + c_3)q_\psi + B_\psi\big],
\]
and hence
\begin{equation}
  q_\psi \ge \frac{c_1 N^2 - B_\psi}{4/\gamma_y + c_2\sqrt{N} + c_3}.
  \label{eq:floor}
\end{equation}
For each fixed vector $\psi$ and fixed $\gamma_y > 0$, the right-hand side grows as $N^{3/2}$, forcing $q_\psi = \infty$. In a fixed-charge lattice sector, \eqref{eq:gmin} gives $4/\gamma_y \le 4s/\pi$, so~\eqref{eq:floor} also holds with $4s/\pi$ in the denominator once $c_1 N^2 > B_\psi$ (uniformity across states or lattices would additionally require uniform remainder constants). The weaker hypothesis $\sum_k |r_k(\psi)| = o(\Lambda_N)$ likewise contradicts~\eqref{eq:cancellation} because $\Lambda_N \to \infty$.
\end{proof}

As noted above, \eqref{eq:HS} motivates Assumption~\ref{ass:O} at leading quadratic order, whereas controlling higher field powers and Moyal corrections at fixed $\hbar$ uniformly in $N$ remains open (Eq.~(C.15e) of~\cite{Ven} controls fixed coefficients only at fixed resolution). Moreover, because $B_\psi$ may depend on $\psi$, inequality~\eqref{eq:floor} applies vector by vector rather than giving a uniform spectral floor.

To translate Theorem~\ref{thm:non-linear} to the corner-weighted clock~\eqref{eq:clock_form}, fix a nonzero vector $\psi$ in the natural corner domain, set $\phi_y = \hat{\varrho}_S(y)\hat{a}^{-1/2}\psi$, and hold these corner factors fixed as the radial regulator is removed. At fixed positive total area, $\int_S \|\phi_y\|^2 \epsilon_0 = \mathcal{A}\|\hat{a}^{-1/2}\psi\|^2 > 0$. If the regulated hypotheses of Theorem~\ref{thm:non-linear} hold on the coupled sector for almost every $y$ and $v$ on a common radial interval, then on each nonzero $\phi_y$ there is no subsequence of bounded focusing expectations. Applying scalar Jensen to $\phi_y/\|\phi_y\|$ at finite resolution followed by Fatou's lemma in~\eqref{eq:clock_form} shows that the clock form diverges on every such fixed $\psi$, so its pointwise limiting form has trivial domain. Identifying this pointwise limit with a resolvent limit or with limits of regulated spectral projections requires a separate convergence argument; in the linearized theory, Corollary~\ref{cor:clock} provides the stronger, state-independent spectral bound.

\section{Normal ordering and angular smearing of the focusing}\label{sec:repairs}

Section~\ref{sec:subtracted-clock} examines vacuum subtraction applied after expanding the affine delay to quadratic order. An alternative repair is to normal-order or smear the focusing operator $\hat{\Xi}_y(v)$ directly inside the exponential of~\eqref{eq:clock_form}. Here, we record why standard quadratic-field prescriptions fail to preserve both a uniform area ceiling and finite dimensionality.

Any modification of $\hat{\Xi}_y(v)$ must still supply a lower bound if it is to cap the corner area. A uniform bound $\hat{\Xi}_y(v) \ge -C\mathbf{1}$, independent of endpoint, area, and resolution, would preserve the area ceiling~\eqref{eq:area_bound} up to a factor $e^{2C}$. For a two-dimensional chiral free field, smooth sampling yields Flanagan's quantum energy inequality $\int g^2 \langle {:}T{:} \rangle \ge -(12\pi)^{-1}\int (g')^2$ in standard normalization~\cite{Flanagan}, whose lower bound deteriorates to $-\infty$ as the window approaches a sharp endpoint. In four dimensions, Fewster and Roman showed that a free scalar field admits states with arbitrarily negative null-geodesic averages even for smooth weights~\cite{FewsterRoman}. Applying these bounds directly to the reduced gravitational shear requires identifying its radial operator and reference state; note that~\cite{Ven} normal-orders only the separately smeared shear in Eqs.~(5.36a) and~(D.28), while retaining the unsubtracted positive ordering in Eq.~(4.22).

Smearing the squared shear \emph{after} normal ordering does not cure the angular instability in the quadratic theory, because it still sums over independent harmonics. Consider identical Gaussian copies labeled by angular harmonics, a smooth interior radial window, and a nonnegative angular kernel of unit integral. Suppose a single copy admits a centered state with negative expectation $-d < 0$ for the sampled square; any nonzero vacuum-to-two-particle matrix element for the chosen window suffices, because a small two-particle admixture of suitable phase interferes negatively at first order. Placing this state in every harmonic through $J$ eliminates cross terms by centering, and the spin-weighted addition theorem gives total expectation $-d\sum_{j=2}^J (2j+1)/(4\pi)$, which diverges to $-\infty$ as $J \to \infty$ and precludes a uniform lower bound. This is the gravitational counterpart of the result that smeared null energy cannot be bounded below by smearing on a single light sheet alone~\cite{FlissFreivogel}; in four-dimensional quantum field theory, a finite lower bound requires smearing over two null directions~\cite{DSNEC}, which is unavailable to an operator built purely from incoming null-cone data.

Smearing the field \emph{before} squaring reproduces the band-limited example of Section~\ref{sec:hidden}. A kernel with an exact angular cutoff omits all harmonics $j > J$ from the clock while leaving their Fock states in $\mathcal{K}_{\mathcal{A}}$, making the projected space manifestly infinite-dimensional (mere suppression of high harmonics does not give this exact factorization). Both remarks apply to quadratic fields; non-linear focusing and corner weights can couple distinct harmonics.

\section{Packet construction and proofs for the quadratic clock}\label{app:packets}

This appendix provides the technical details for Section~\ref{sec:subtracted-clock}. We construct the localized radiative packets satisfying the linearized endpoint conditions, prove Theorem~\ref{thm:subtracted-clock}, and evaluate the exact classical clock reading on the same profiles to establish the amplitude bound~\eqref{eq:scaled-pairing}.

\subsection{Displacements of arbitrarily small quadratic delay}

Work at fixed positive area on a round, shear-free background of radius $R = \sqrt{\mathcal{A}/(4\pi)}$, where the background clock reads $(1 - \epsilon)R$. In the radiative tangent space, restrict to a single real tensor harmonic of degree $j \ge 2$ and one polarization, with radial profile $h(v)$. Expanding the averaged affine delay~\eqref{eq:classical_clock} through second order before quantizing gives $e^{\Xi_y} = 1 + \Xi_y^{[2]} + O(z^3)$, where $\Xi_y^{[2]}$ is the term quadratic in the radiative perturbation. Interchanging the radial integrals yields the classical quadratic delay $q(h)$, while Eq.~(D.42a) of~\cite{Ven} gives the radiative symplectic pairing $\Omega(h, g)$:
\begin{align}
  q(h) &= a\int_\epsilon^1 v(1 - v)|h'(v)|^2\,dv, \qquad a > 0, \nonumber\\
  \Omega(h, g) &= b\int_\epsilon^1 v^2\bigl(h'(v)g(v) - g'(v)h(v)\bigr)\,dv, \qquad b \neq 0,
  \label{eq:delay-pairing}
\end{align}
where the constants $a, b$ absorb the harmonic normalization. Here $q(h)$ denotes the classical delay increment of a displacement, distinct from the focusing expectation $q_\psi$ of Section~\ref{sec:weyl}. The factor $1 - v$ in $q(h)$ measures the remaining radial interval over which shear at $v$ delays the generator (in contrast to the stronger quadratic form in Eq.~(D.40) of~\cite{Ven}, which includes transverse focusing and has a non-vanishing weight at $v = 1$).

To ensure the displacements are compatible with the linearized endpoint data, we choose smooth packets compactly supported in $(\epsilon, 1)$, so that their conformal metric variations vanish near both cuts while the area data, angular identification, and affine normalization at the cap remain fixed. Around zero shear, the linear variation of the focusing vanishes, so~\eqref{eq:classical_clock} gives zero first-order shift of the affine endpoint (see Eq.~(A.22) of~\cite{Ven}). Matching the normal connection to the unperturbed opposite sheet imposes one additional linear condition. With zero cap variation, integrating the linearized angular constraint (Eq.~(D.39) of~\cite{Ven}) yields
\[
  \delta\varpi_I(1) = -\frac{t_j}{2}\int_\epsilon^1 v^2 h'(v)\,dv
  = t_j \int_\epsilon^1 v h(v)\,dv,
  \qquad t_j^2 = \frac{(j-1)(j+2)}{2}.
\]
Compact support alone does not force this moment to vanish, so we impose $\int_\epsilon^1 v h(v)\,dv = 0$ on both packets. The normal-connection variation then vanishes near both cuts, and the area and normal-connection matching of Eq.~(A.29) of~\cite{Ven} hold to first order with the opposite sheet and relative boost fixed.

To satisfy this moment condition without losing the symplectic pairing, change variables to $t = -\log(1 - v)$. In this coordinate, $q(h) = a\int v|\partial_t h|^2\,dt$, while $\Omega(h, g) = b\int v^2[(\partial_t h)g - (\partial_t g)h]\,dt$. Choose a nonzero real bump $F \in C_c^\infty(0,1)$ and set $G = F'$, so that $\int_0^1 (F'G - G'F) = 2\int_0^1 (F')^2 > 0$. For $L > 0$ and $i = 1, 2, 3$, define the disjointly supported shifted profiles $F_i(t) = F((t - 2iL)/L)$ and $G_i(t) = G((t - 2iL)/L)$. Pick a real unit vector $(c_1, c_2, c_3) \in \mathbb{R}^3$ orthogonal to both moment vectors $(\int v F_i\,dv)_{i=1}^3$ and $(\int v G_i\,dv)_{i=1}^3$ (with the profiles evaluated at $t(v)$). Then $h_L = \sum_{i=1}^3 c_i F_i$ and $g_L = \sum_{i=1}^3 c_i G_i$ satisfy the endpoint moment condition exactly for every sufficiently large $L$. Because $v \to 1$ uniformly on the support of each packet as $L \to \infty$, disjoint support and $\sum_{i=1}^3 c_i^2 = 1$ give
\begin{equation}
  q(h_L) + q(g_L) = O(L^{-1}), \qquad
  \Omega(h_L, g_L) \longrightarrow 2b\int_0^1 (F')^2 \neq 0.
  \label{eq:soft-packets}
\end{equation}
If the linearized boundary problem fixes additional transverse data, Eqs.~(3.57c)--(3.57e) of~\cite{Ven} transport the connection, expansion, and shear along the null ray; on a spherical background, a fixed harmonic contributes only finitely many endpoint components. For $m$ homogeneous linear endpoint conditions in total, replacing the three intervals by $2m + 1$ disjoint intervals and choosing a unit coefficient vector in the kernel of the resulting $2m$ linear constraints leaves~\eqref{eq:soft-packets} unchanged. These profiles are linearized tangent perturbations; extending them to exact non-linear solutions with all cap data held fixed is a separate boundary-value problem.

To connect~\eqref{eq:soft-packets} to a vacuum-subtracted quantum clock, let $W(f)$ denote the Weyl unitary implementing a classical displacement by $f$ on the linearized field, and let $u[\psi]$ denote the quadratic form of the subtracted clock. Under ordinary point splitting of the quadratic delay, one subtracts a state-independent two-point kernel before taking coincidence. For a shear component $J$, the sum of the products with $J$ shifted by $+f'$ and $-f'$, minus twice the unshifted product, equals $2f'(v)f'(v')\mathbf{1}$ at separated points, so the state-independent subtraction cancels identically. Taking coincidence and integrating yields the displacement identity
\begin{equation}
  u[W(f)\psi] + u[W(-f)\psi] - 2u[\psi] = 2q(f)\|\psi\|^2.
  \label{eq:displacement-delay}
\end{equation}
Applying~\eqref{eq:displacement-delay} on spectral subspaces requires that it hold on the continuum form domain of the clock. A sufficient criterion is convergence on a common Weyl-invariant form core to a closable, lower-bounded form, with the separated-point identity passing to core values. After shifting the lower bound to zero, \eqref{eq:displacement-delay} bounds the form norm of $W(\pm f)\psi$ by a constant multiple of the form norm of $\psi$, so core approximation extends both the displacements and~\eqref{eq:displacement-delay} to the closed form domain. Classical admissibility of a smooth packet does not by itself establish this quantum criterion.

\begin{proof}[Proof of Theorem~\ref{thm:subtracted-clock}]
Suppose for contradiction that $\mathbf{1}_{(-\infty, \ell]}(U)$ has finite rank for some $\ell > \inf\sigma(U)$. Then $E_0 = \inf\sigma(U)$ is an eigenvalue with a finite-dimensional eigenspace $P_0\mathcal{H}$ separated by a positive spectral gap $\Delta > 0$ from its orthogonal complement. For any unit vector $\psi \in P_0\mathcal{H}$, both increases $u[W(\pm f)\psi] - E_0$ in~\eqref{eq:displacement-delay} are nonnegative, so the gap implies
\[
  \|(1 - P_0)W(\pm f)P_0\|^2 \le \frac{2q(f)}{\Delta}.
\]
By~\eqref{eq:soft-packets}, we may rescale every $g_L$ by a single fixed real factor so that the limiting Weyl phase $e^{i\theta}$ has $\theta/(2\pi)$ irrational, where $\theta = \pm\lim_{L\to\infty}\Omega(h_L, g_L)/\hbar$ depending on orientation, while keeping $q(h_L) + q(g_L) \to 0$. Compressing $W(h_L)$ and $W(g_L)$ to the finite-dimensional space $P_0\mathcal{H}$, the gap estimate ensures that their compressions have a common subsequence converging to unitary operators $A, B$ on $P_0\mathcal{H}$, and that the error in replacing a compressed product by the product of the compressions vanishes in the limit. The Weyl relation therefore descends to $AB = e^{i\theta}BA$ on $P_0\mathcal{H}$. Taking determinants gives $e^{i\theta\dim(P_0\mathcal{H})} = 1$, contradicting the irrationality of $\theta/(2\pi)$.
\end{proof}

\subsection{Exact classical readings on the same packets}

Because the packets~\eqref{eq:soft-packets} achieve vanishing quadratic delay while their radial derivatives grow, we test the uniformity of the quadratic expansion by evaluating the exact classical affine time~\eqref{eq:classical_clock} on incoming data with the same radial profiles. Hold the round corner fixed and give $h_L$ a real amplitude $\alpha_L$. Let $T(y)$ be the trace-free endomorphism obtained by raising one index of the tensor harmonic with the round conformal metric $\bar{q}_0$. For the incoming conformal metric $\bar{q}(v,y) = \bar{q}_0(y)\exp[\alpha_L h_L(v)T(y)]$, we have $\bar{q}^{-1}\partial_v\bar{q} = \alpha_L h'_L T$ exactly. Denoting the normalized spherical average by $\langle\cdot\rangle_S$, equation~\eqref{eq:classical_xi} gives, after the packet has passed,
\[
  \langle\Xi_y\rangle_S
  = \frac{\alpha_L^2}{8}\langle\operatorname{tr}(T^2)\rangle_S \int_\epsilon^1 v|h'_L(v)|^2\,dv.
\]
On the three packet intervals ($i = 1, 2, 3$), we have $1 - v \le e^{-2L}$, while $Lq(h_L) \to a\int_0^1 |F'|^2 > 0$; hence $\int_\epsilon^1 v|h'_L|^2\,dv$ is bounded below by a positive constant times $e^{2L}/L$. The third interval ends at $t = 7L$, corresponding to $v_L = 1 - e^{-7L}$, and the accumulated focusing persists from $v_L$ to the corner $v = 1$. Restricting~\eqref{eq:classical_clock} to $[v_L, 1]$ and applying Jensen's inequality on $S$ yields
\begin{equation}
  \bar{u}_{*,L}
  \ge R(1 - v_L)\exp\!\bigl[\langle\Xi_y(v_L)\rangle_S\bigr]
  \ge R\exp\!\left(-7L + c\,\alpha_L^2 e^{2L}/L\right),
  \label{eq:exact-packet-delay}
\end{equation}
with $c > 0$ independent of $L$ and $\alpha_L$. Every fixed nonzero amplitude therefore yields a diverging exact classical reading as $L \to \infty$.

For a fixed threshold $\ell > 0$, keeping $\bar{u}_{*,L} \le \ell$ requires
\[
  |\alpha_L|^2 \le c^{-1} L e^{-2L}\bigl[7L + \log(\ell/R)\bigr].
\]
Repeating the same calculation for $g_L$ (replacing $F$ by $G$) imposes the same order of bound on its amplitude $\beta_L$. Thus, if both exact readings remain below $\ell$, both amplitudes are $O(L e^{-L})$. By bilinearity of the tangent-space symplectic form and~\eqref{eq:soft-packets},
\begin{equation}
  |\Omega(\alpha_L h_L, \beta_L g_L)| = O(L^2 e^{-2L}) \longrightarrow 0.
  \label{eq:scaled-pairing}
\end{equation}
At fixed $\hbar$, the Weyl phase used in Theorem~\ref{thm:subtracted-clock} therefore tends to unity on these rescaled displacements. This calculation tests exact incoming profiles at fixed corner area with matching to the opposite sheet checked to first order; the reading of a renormalized exact quantum clock on displaced states remains open.

Two comparisons with Appendix~D.3 of~\cite{Ven} situate this behavior. First, the radial shear term in the stronger quadratic form of Eq.~(D.40) carries weight $v^2/8$, which stays positive at $v = 1$; the same estimate on $\int_\epsilon^1 v|h'_L|^2\,dv$ shows that Eq.~(D.40) grows at least as $e^{2L}/L$ on unit-amplitude packets. Second, Eqs.~(D.50)--(D.55) bound the symplectic area enclosed by a loop of four classical profiles confined to a shrinking \emph{angular} patch at bounded exact reading and fixed corner data. Our packets occupy a single fixed angular harmonic and approach the corner \emph{radially}, and~\eqref{eq:scaled-pairing} exhibits how the exact reading constrains that radial construction.

\end{document}